\documentclass{PoS}

\title{Search for Dark Matter in pp Collisions with CMS}

\ShortTitle{Search for Dark Matter in pp Collisions with CMS}

\author{\speaker{Shin-Shan~Yu}\\ 
        On behalf of the CMS Collaboration \\
        Department of Physics, National Central University, Chung-Li, Taiwan\\
        E-mail: \email{syu@cern.ch}}

\abstract{
Searches in CMS for dark matter in final states with invisible particles 
recoiling against jets, top, W, Z, photon, and Higgs are presented. Various 
topologies are explored, covering several specific dark-matter production 
modes. The summary in a simplified-model framework of various searches 
for direct dark matter production with the CMS detector is discussed, 
highlighting sensitivities of the analyses under various assumptions of DM 
production.
}

\FullConference{38th International Conference on High Energy Physics\\
		3-10 August 2016\\
		Chicago, USA}

\newcommand{\mmed}{\ensuremath{M_{\mathrm{Med}}}}
\newcommand{\mdm}{\ensuremath{m_{\mathrm{DM}}}}

\newcommand{\pt}{\ensuremath{p_{\mathrm{T}}}}

\newcommand{\ttbar} {\ensuremath{\mathrm{t}\bar{\mathrm{t}}}}
\newcommand{\fbinv} {\ensuremath{\mathrm{fb}^\mathrm{-1}}}
\newcommand{\ETm}{\ensuremath{E_{\mathrm{T}}^{\mathrm{miss}}}}
\newcommand{\ptvecmiss}{\ensuremath{{\vec p}_{\mathrm{T}}^{\kern1pt\mathrm{miss}}}}

\begin{document}

\section{Introduction}
Although the dominant component of the matter in the universe is dark matter 
(DM), its underlying particle nature remains unknown and cannot be explained 
within the standard model (SM)~\cite{FNAL_Review}. 
If non-gravitational interactions exist between DM and SM 
particles, DM particles could be produced at the CERN LHC. 
The searches for dark matter at LHC, complementary to the direct detection 
experiments via DM-nucleon scattering and to the indirect 
detection experiments via DM annihilation, can probe a wide range of 
DM-SM interaction types. The CMS searches 
at Run II use the benchmark simplified models in Ref.~\cite{DMF} to 
interpret results. These benchmark models are established with the following 
assumptions: (i) DM is a single particle, a Dirac fermion, stable on collider 
timescales and non-interacting with the detector, pair-produced, (ii) there is
 a new massive particle that mediates the DM-SM interaction, (iii) minimal 
flavor violation, i.e. spin-0 mediator must have couplings to fermions 
proportional to the SM Higgs couplings, (iv) mediator has minimal decay width;
only decays strictly necessary for the self-consistency of the model
(e.g. to DM and to quarks) are accounted for in the definition of 
the mediator width. 
In these models, the minimal set of parameters include 
coupling structure, the mediator and DM masses \mmed\ and \mdm, 
and coupling of mediator to SM and DM particles 
$g_\mathrm{q}$ and $g_\mathrm{DM}$. 

One way to observe DM particles at LHC is through their recoil off of SM 
particle X that is produced in association with the DM. The following 
text describes the analysis where X is a jet or a hadronically-decaying 
W or Z boson~\cite{EXO-16-037}, a leptonically-decaying Z~\cite{EXO-16-038}, 
a photon~\cite{EXO-16-039}, a pair of top quarks \ttbar~\cite{EXO-16-005}, or a single top quark~\cite{EXO-16-040}, and 
a Higgs boson~\cite{EXO-16-011,EXO-16-012}. 
The summary of the mono-X searches is also discussed. 
The mono-Higgs and mono-\ttbar\ searches 
were performed based on 2.3~\fbinv\ of 2015 data while the rest of the mono-X 
searches were performed using 12.9~\fbinv\ of 2016 data in $\sqrt{s}=13$~TeV 
pp collisions collected with the CMS detector~\cite{CMS}. 
Overall, if the analysis contains jets 
in the final state, the dominant background contribution comes from
 the \ttbar\ production 
and associated production of jets with an invisibly-decaying Z boson 
(Z$\rightarrow\nu\nu$) or a leptonically-decaying W boson 
(W$\rightarrow\ell\nu$). Otherwise, the dominant background comes from the 
SM diboson or tri-boson production, such as ZZ, W$\gamma$, Z$\gamma$, 
W$\gamma\gamma$, or Z$\gamma\gamma$, where the W boson decays leptonically 
and one of the Z bosons decays invisibly.

\section{Mono-jet/jets/hadronic-W,Z\label{sec:monojet}}
We first pre-select events containing large \ETm\ ($>200$~GeV), 
at least one AK4 jet\footnote{
The clustering of jets at CMS is performed with the anti-kt 
algorithm~\cite{Cacciari:2008gp} with a distance parameter of 0.4 and 0.8 
(denoted as AK4 and AK8 jets), or with the 
Cambridge-Aachen algorithm~\cite{Catani:1993hr} with a distance parameter 
of 1.5 (denoted as CA15 jets), respectively. as implemented in the 
\textsc{FASTJET} package~\cite{fastjet}.}
 with $\pt>100$~GeV and veto events with well-identified 
electrons, muons, taus, photons, and b-jets. 
The minimum azimuthal angle $\Delta\phi$ between the \ptvecmiss\ direction 
and each of the first four leading AK4 jets with \pt\ greater than 30~GeV is 
required to be greater than 0.5. 
Events are further classified into mono-V or mono-jet category. 
An event falls into the mono-V category if $\ETm > 250$~GeV, the leading AK8 
jet in the event has $\pt > 250$~GeV and $|\eta| < 2.4$, the jet 
mass after pruning within 65--105~GeV~\cite{Ellis:2009me}, and the ratio 
of $N$-subjettiness $\tau_{2}/\tau_{1}$ less than 0.6~\cite{Thaler:2010tr}.  
The rest of the pre-selected events fall into the mono-jet category. 
Ten mutually exclusive control regions in data are used in order to get a 
precise estimate of the dominant Z+jets and W+jets backgrounds: 
dimuon, dielectron, single-muon, single-electron, and $\gamma$+jets events 
that satisfy requirements resembling the selections imposed on the 
mono-V and mono-jet categories. The \ETm\ in these control regions is 
redefined by excluding the leptons or the photon from the \ETm\ calculation. 
The resulting hadronic recoil mimics the \ETm\ shape of the backgrounds in 
the signal region. 
Transfer factors that take into account the difference between 
the signal region and the control regions in the differential 
cross section, branching ratio, acceptance, reconstruction and identification 
efficiencies are derived from simulation; whenever possible, \pt-dependent 
NLO QCD and NLO electroweak $K$ factors extracted from theoretical 
calculations are applied. The shape and the normalization of the 
\ETm\ spectra for the dominant background is determined through a maximum 
likelihood fit, performed simultaneously across all \ETm\ bins 
in the ten control regions and the two signal regions. 
Figure~\ref{fig:MET} shows the \ETm\ distributions in the mono-jet and mono-V 
signal regions.
No significant excess is observed with respect to the SM backgrounds. 
Limits are computed on the DM production cross section using simplified 
models with spin-1 or spin-0 mediators. 
Vector and axial-vector mediators with masses up to 1.95~TeV are excluded at 
95\% CL. Scalar and pseudoscalar mediators with masses up to 100 and 430~GeV, 
respectively, are excluded at 95\% CL. The search yields an observed 
(expected) upper limit of 0.44 (0.56) at 95\% CL on the invisible branching 
fraction of the 125 GeV Higgs boson assuming SM production cross section. 

\section{Mono-leptonic-Z\label{sec:monoz}}
We select events with \ETm\ $>100$~GeV, containing a pair of electrons or 
muons with $\pt^{\ell\ell}>60$~GeV and reject events with extra 
well-identified electrons, muons, taus, b-jets, and events with more than one 
jet. The dominant background from the ZZ/WZ diboson production is estimated 
with simulated events including NNLO QCD and NLO EWK corrections. The 
background from \ttbar, W+jets, WW, tW, and Z$\rightarrow\tau\tau$ is 
estimated 
from the $e\mu$ data with a correction factor of $0.5\sqrt{N_{ee(\mu\mu)}^\mathrm{data}/N_{\mu\mu(ee)}^\mathrm{data}}$ for the $ee$ ($\mu\mu$) channel; the 
factor 0.5 corrects for the branching ratio difference while the ratio 
$\sqrt{N_{ee}^\mathrm{data}/N_{\mu\mu}^\mathrm{data}}$ 
accounts for the difference in the reconstruction and identification 
efficiency between electrons and muons. 
Results are interpreted with DM models with 
vector/axial-vector mediators and the invisible decays of the Higgs boson. 
Assuming the SM production rate, the observed (expected) 95\% CL upper limit 
on ${\cal B}(\mathrm{H}\rightarrow \mathrm{inv.})$ is 0.86 (0.70).

\section{Mono-photon\label{sec:monogamma}}
We select events containing large \ETm\ ($>170$~GeV), 
at least one photon with $\pt>175$~GeV and $|\eta|<1.44$, and veto events with 
well-identified electrons and muons.  
The minimum $\Delta \phi$ requirement in Section~\ref{sec:monojet} 
is also applied to reduce the QCD background. 
The major background from the Z$(\rightarrow \nu\nu)\gamma$ and 
W$(\rightarrow \ell \nu) \gamma$ processes 
is estimated using simulated events with NNLO QCD and NLO EWK corrections and 
cross-checked with control data dominated by well-reconstructed 
Z$(\rightarrow \ell^+\ell^-)\gamma$ and W$(\rightarrow \ell \nu) \gamma$ 
events. Background from jets or electrons mis-identified as photons is 
estimated by measuring the mis-identification rates in control samples in 
data. Non-collision background from beam halo and the direct 
interaction of particles with the ECAL photodetector are estimated by fits 
to distributions of the photon $\phi$ and the EM cluster seed time: 
beam halo events tend to produce photons with $\phi \sim 0, \pi$, while the 
rest of the photon events tend to be uniformly distributed in $\phi$; each 
process also exhibits a distinctive distribution in the EM cluster seed time 
and one could fit the distribution in data to the templates to extract 
the contribution of each component. The number of events observed in data 
is in good agreement with the total expected background. For the simplified 
DM models considered, vector/axial-vector mediator masses of up to 760~GeV are 
excluded for small \mdm. The suppression scale in the effected 
field theory (EFT, dimension-7) $\Lambda$ is excluded at 95\% CL up to 
620~GeV. The true scale of the gravitational interaction in the ADD extra 
dimension model is excluded from below 2.44 to 2.60 TeV for n=3--6 extra 
dimensions. 

\section{Mono-\ttbar \label{sec:monott}}
The analysis incorporates both hadronic and semileptonic \ttbar\ 
final states in a combined search. We require $\ETm>200~(160)$~GeV for the 
 hadronic (semileptonic) channel. The major background 
comes from \ttbar\ production with one less hadronic top; i.e. 
semileptonic (full-leptonic) \ttbar\ events for the hadronic (semileptonic) 
 mono-\ttbar\ search. A novel resolved-hadronic-top-tagging technique 
is developed, combining the information of the quark/gluon discriminant value 
for each jet, values of the b-tag discriminants, opening angles between the 
candidate b jet and each of the jets from the candidate W boson, and 
the $\chi^{2}$ of a simultaneous kinematic fit to the top quark and W bosons 
masses using the reconstructed jet momenta, energy, and resolutions. 
Sensitivity is further improved by up to 30\% after categorizing hadronic channel 
events by expected signal purity based on the number of top tags, b-tagged jets, 
and $\Delta \phi(jet,\ptvecmiss)$. The search is interpreted in terms of DM 
production to place constraints on the parameter space of simplified models with 
spin-0 mediators.

\section{Mono-top\label{sec:monot}}
In this search we consider events with $\ETm>250$~GeV 
and a hadronically-decaying top quark reconstructed using a CA15 jet.
Weights calculated with the PileUp Per Particle Identification (PUPPI) 
algorithm~\cite{puppi} are applied to the particle-flow candidates to account for 
the impact of pileups. The CA15 jet must have $\pt>250$~GeV, a softdrop mass within 
110--210~GeV~\cite{msd} and the ratio of $N$-subjettiness $\tau_{3}/\tau_{2}$ less 
than 0.61. Dominant background from \ttbar, Z+jets, and W+jets 
 is estimated using constraints from seven control regions:
 dimuon, dielectron, $\gamma$+jets, single muon ($b$-tagged and anti-$b$-tagged), 
 and single electron ($b$-tagged and anti-$b$-tagged). 
Similar to Section~\ref{sec:monojet}, the shape and the normalization of the \ETm\ 
spectra is determined through a maximum likelihood fit, performed simultaneously 
across all \ETm\ bins in the control regions and the signal regions. 
Results are interpreted in terms of DM produced via a neutral flavor-changing 
interaction or via the decay of a colored, scalar resonance together with a single top 
quark. For the non-resonant model, assuming $\mdm = 10$~GeV and $a_\mathrm{FC}=b_\mathrm{FC}=0.25$, flavor-changing neutral currents of $\mmed<1.5$~TeV are 
excluded at 95\% CL. For the resonant model, scalar fields with 
$\mmed<2.7$~TeV are excluded at 95\% CL.

\section{Mono-Higgs \label{sec:monoh}}
Because Higgs boson radiation from an initial-state quark is 
Yukawa-suppressed and in a potential signal the Higgs boson would be part of 
the interaction producing the DM, mono-Higgs searches have a uniquely 
enhanced sensitivity to the structure of DM-SM couplings. 
We search for DM in the mono-Higgs channel
in which the Higgs boson decays to either a pair of bottom quarks 
(b$\bar{\mathrm{b}}$) or a pair of photons ($\gamma\gamma$). 
The results have been interpreted using a
two-Higgs-doublet model, where a vector boson Z' is 
produced resonantly and decays into the 125 GeV Higgs boson and an 
intermediate heavy pseudoscalar particle A$^0$, which in turn decays into 
two DM particles. The minimum angular distance 
between the decay products of the Higgs boson follows the
relation $\Delta R \approx 2 \times m_\mathrm{h}/p_\mathrm{h}$, where 
$p_\mathrm{h}$ is the momentum of the Higgs boson and increases with the 
mass of Z'. Therefore, the analysis in the bb channel is divided into two 
regimes: (i) a resolved regime where the Higgs boson decays to two distinctly 
reconstructed AK4 b jets, and (ii) a boosted regime where the Higgs boson is 
reconstructed by one single AK8 jet.
The signal extraction is performed with a simultaneous fit to the 
\ETm\ distributions (three bins from 170--1000~GeV for the resolved 
and 200--1000~GeV for the boosted regime) in the signal and 
background-enriched control regions. 
The search in the $\gamma\gamma$ channel is performed by looking for an 
excess in the diphoton mass spectrum after requiring 
$\ETm>105$~GeV. Data driven techniques are used to estimate the reducible 
backgrounds which mainly consists of diphoton SM production. A cut-and-count 
based approach is used to determine the signal yield. 
The Z' mass range of 600 to 1863 GeV is excluded with 95\% CL, 
assuming the coupling parameter $g_\mathrm{Z'} = 0.8$ for A$^0$ mass at 
300~GeV.

\section{Summary of dark matter searches at CMS and conclusion\label{sec:summary}}
\begin{figure}

\begin{center}
\begin{tabular}{cc}
\includegraphics[width=.4\textwidth]{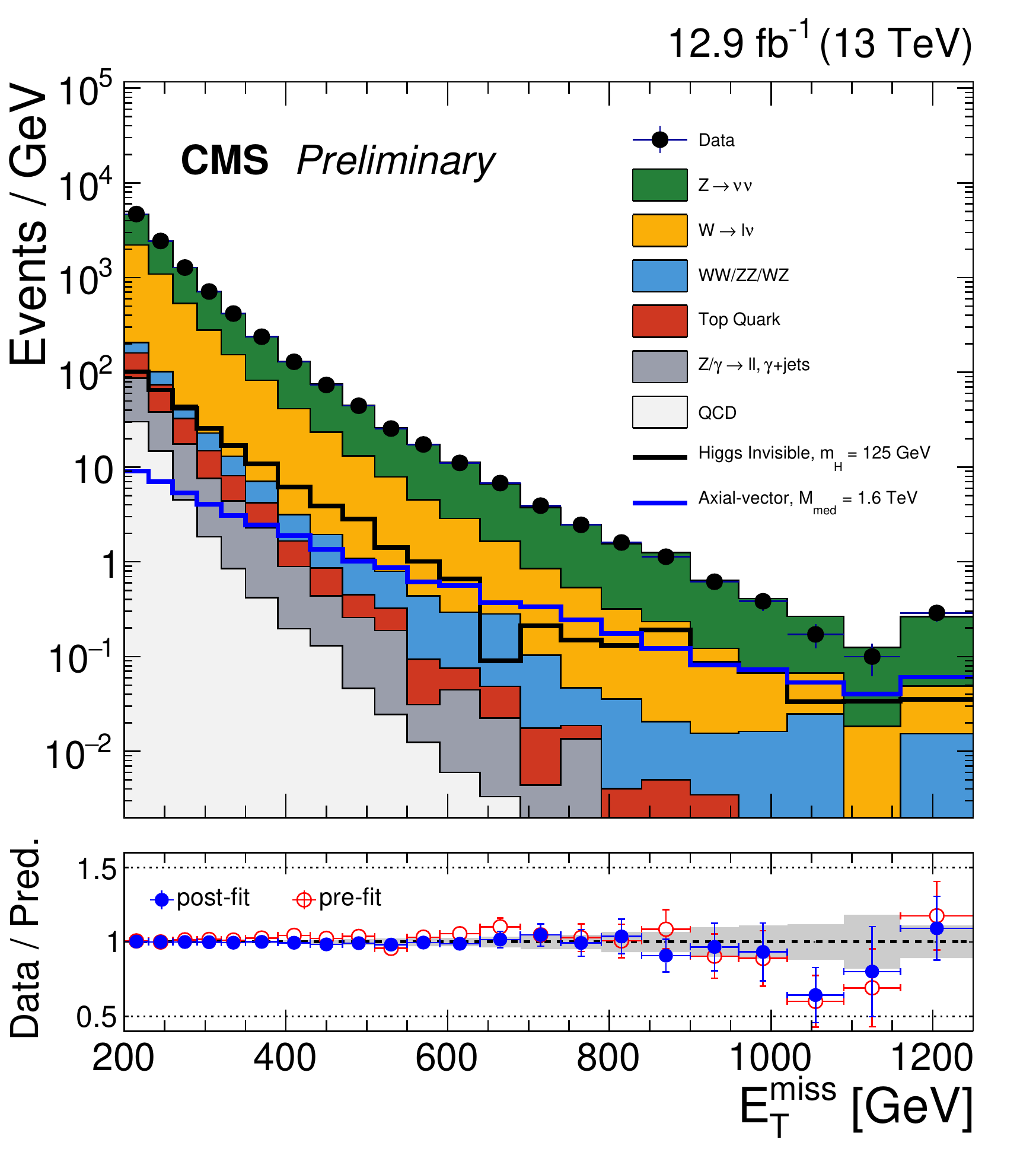}  &
\includegraphics[width=.4\textwidth]{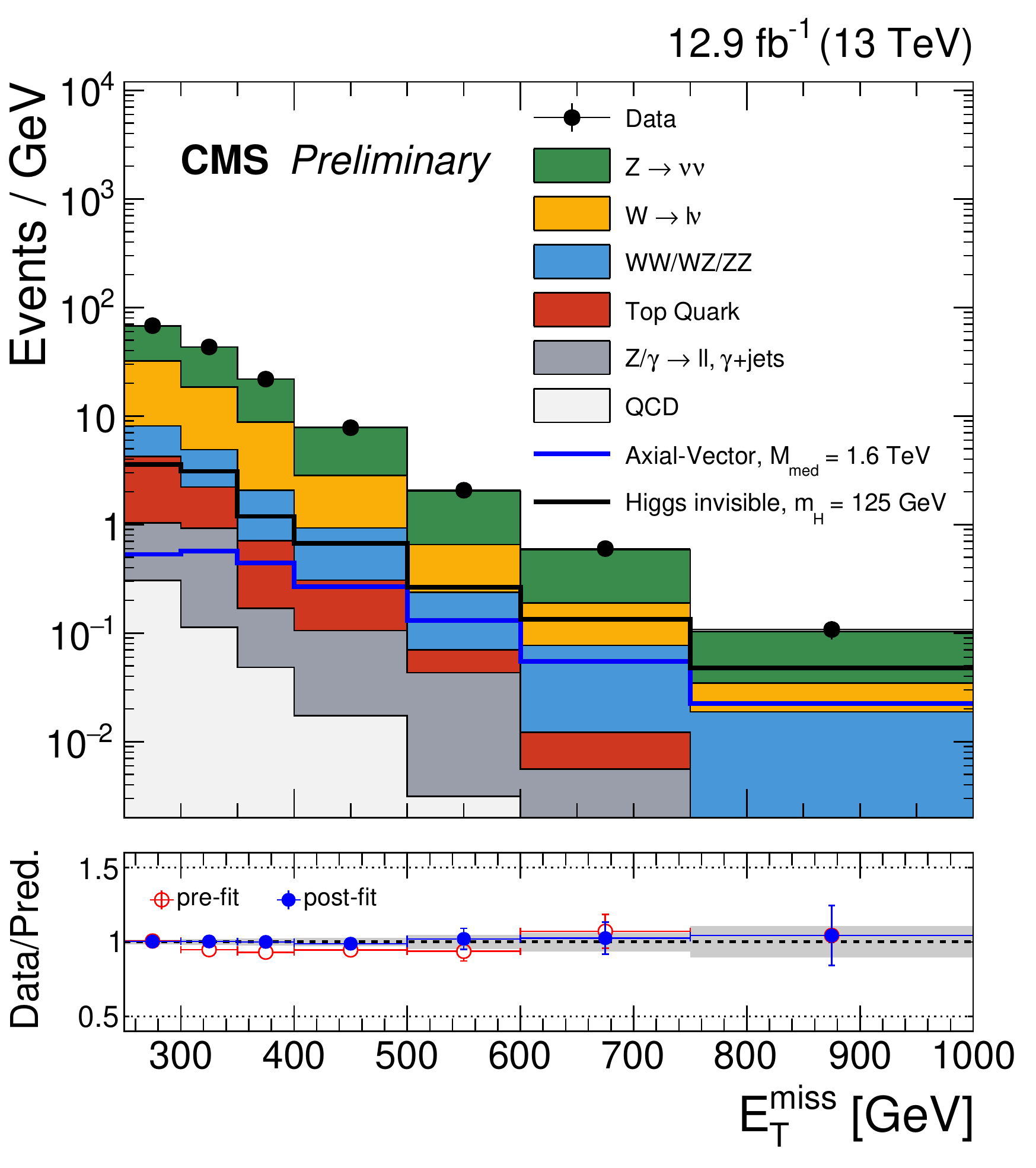}\\
\end{tabular}
\caption{Observed \ETm\ distribution in the monojet (left) and mono-V (right) 
signal regions compared with the post-fit background expectations for various SM processes. \label{fig:MET}}
\begin{tabular}{cc}
\includegraphics[width=.55\textwidth]{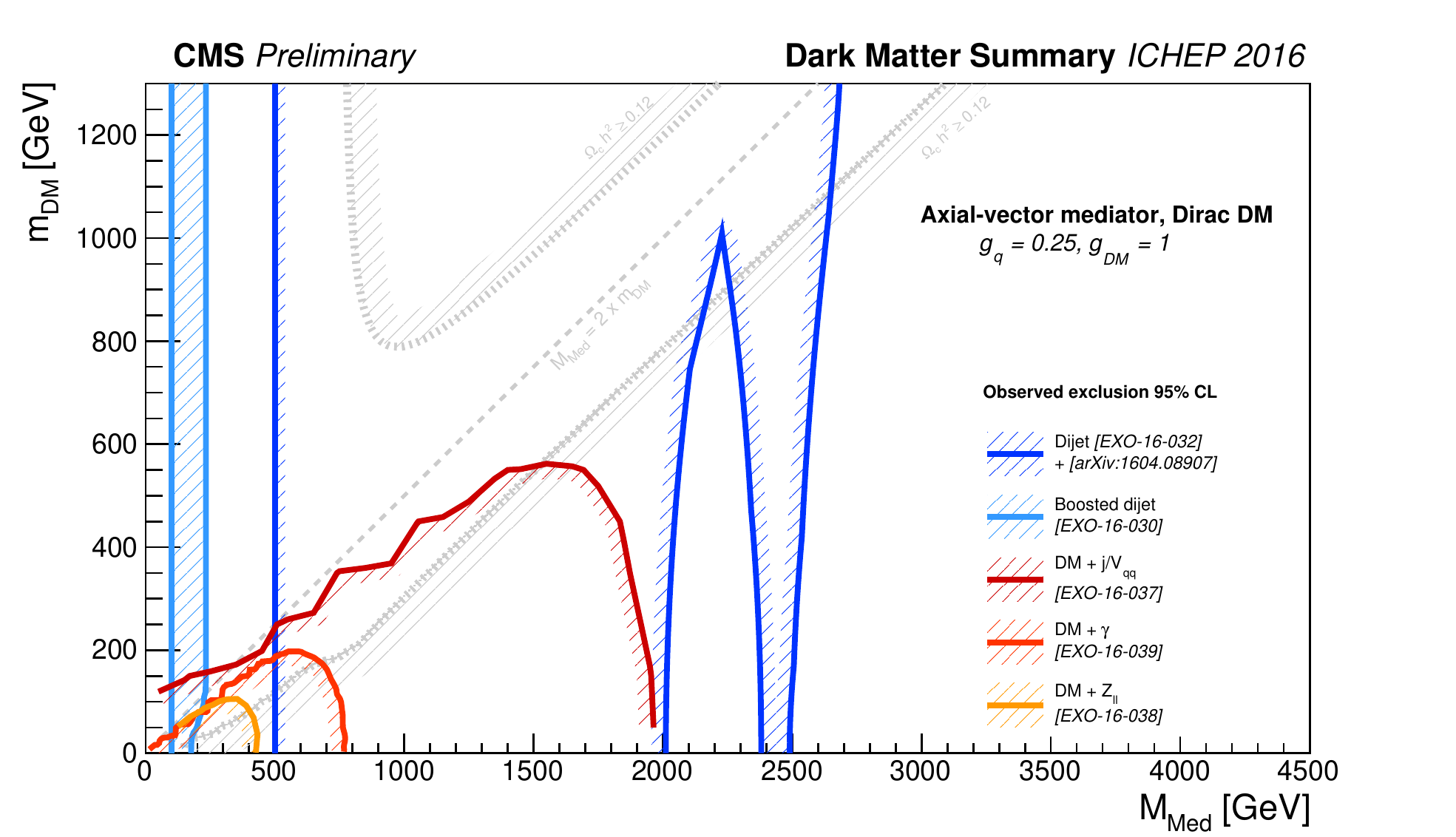}  &
\includegraphics[width=.45\textwidth]{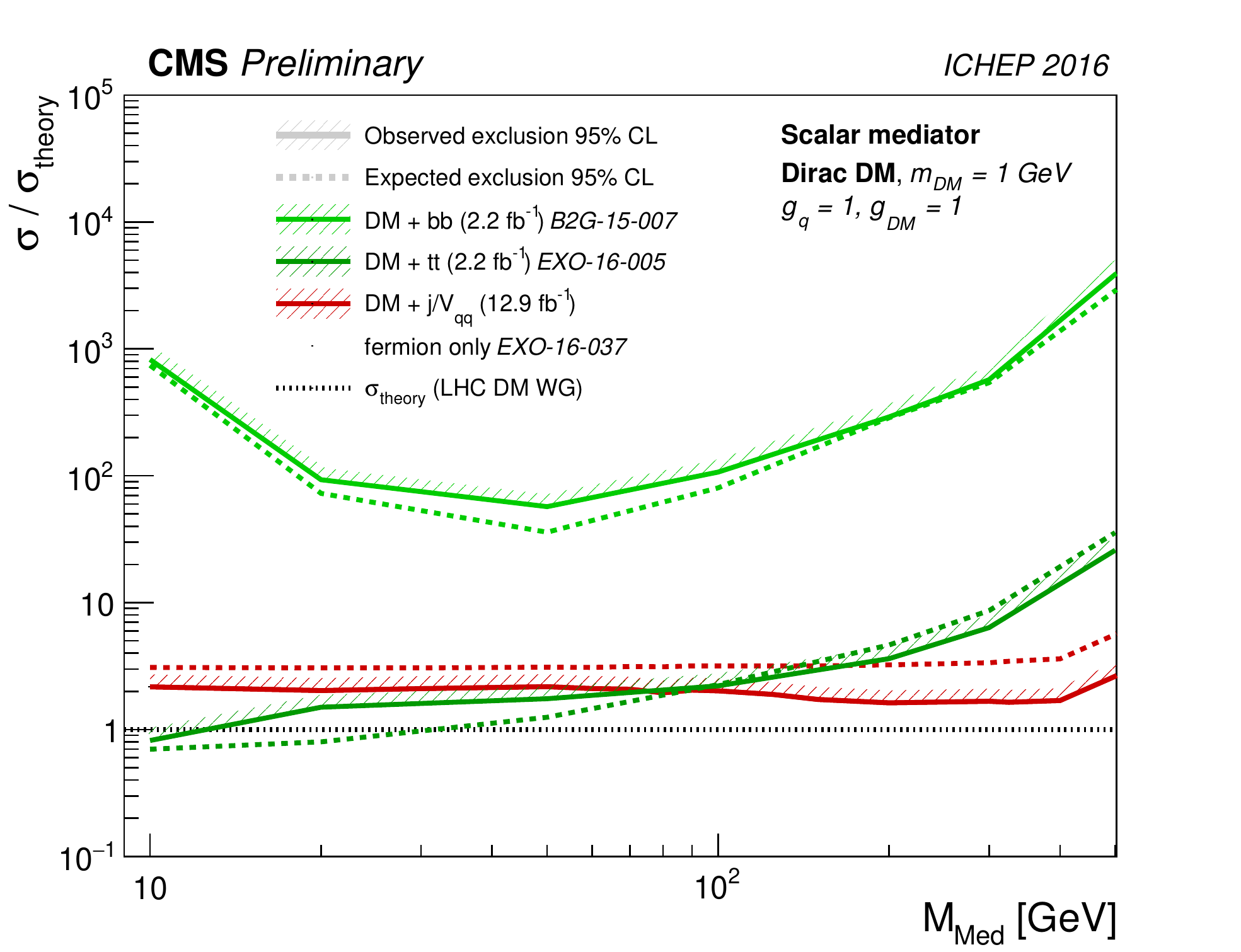}\\
\end{tabular}
\end{center}
\caption{Left: The 95\% CL exclusion regions in $\mmed-\mdm$ plane for di-jet searches 
 and various \ETm\ based DM searches from CMS in the leptophobic axial vector 
 model. Right: The 95\% CL observed (full-line) and expected (dashed-line) 
 exclusion limits for the scalar model as a function of \mmed\ for different 
 \ETm\ based DM searches from CMS.\label{fig:summary}}

\end{figure}

Figure~\ref{fig:summary} (left) shows the 95\% CL exclusion region in the 
$\mmed-\mdm$ plane for di-jet searches and mono-X searches, interpreted 
using a DM model with a leptophobic axial vector mediator, assuming 
$g_\mathrm{q}=0.25$ and $g_\mathrm{DM}=1$; the dijet searches are complimentary 
to the mono-X searches and cover the off-shell region that mono-X is 
less sensitive to. Figure~\ref{fig:summary} (right) shows the exclusion limits 
for the scalar model as a function of \mmed\ from various mono-X searches,
assuming $g_\mathrm{q}=1$ and $g_\mathrm{DM}=1$; for smaller \mmed, the 
mono-\ttbar\ search already has better sensitivity than the mono-jet search 
even with the 2015 dataset. 
We have performed searches for dark matter with various mono-X final states, 
using 2.3~\fbinv\ of 2015 data and 12.9~\fbinv\ of 2016 data in $\sqrt{s}=13$~TeV pp 
collisions collected with the CMS detector. The data are found to be in agreement with 
the SM prediction. We expect updates with the full 2016 dataset in the near future.

\begin{thebibliography}{99}


\bibitem{FNAL_Review} 
  G.~Bertone, D.~Hooper, and J.~Silk,
  ``Particle dark matter: Evidence, candidates and constraints,''
  Phys.\ Rept.\  {\bf 405}, 279 (2005).

\bibitem{DMF} 
  D.~Abercrombie {\it et al.},
  ``Dark Matter Benchmark Models for Early LHC Run-2 Searches: Report of the ATLAS/CMS Dark Matter Forum,''
  arXiv:1507.00966.

\bibitem{EXO-16-037} 
  CMS Collaboration,
  ``Search for dark matter in final states with an energetic jet, or a hadronically decaying W or Z boson using 12.9 fb$^{-1}$ of data at $\sqrt{s}$=13 TeV,''
  CMS-PAS-EXO-16-037.

\bibitem{EXO-16-038} 
  CMS Collaboration,
  ``Search for dark matter in Z+ \ETm\ events using 12.9 fb$^\mathrm{-1}$ of 2016 data,''
  CMS-PAS-EXO-16-038.

\bibitem{EXO-16-039} 
  CMS Collaboration,
  ``Search for dark matter and graviton produced in association with a photon in pp collisions at $\sqrt{s}=13$~TeV with an integrated luminosity of 12.9 fb$^{-1}$,''
  CMS-PAS-EXO-16-039.


\bibitem{EXO-16-005} 
  CMS Collaboration,
  ``Search for dark matter in association with a top quark pair at $\sqrt{s}$=13 TeV,''
  CMS-PAS-EXO-16-005.

\bibitem{EXO-16-040} 
  CMS Collaboration,
  ``Search for new physics in a boosted hadronic monotop final state using 12.9 fb$^{-1}$ of $\sqrt{s}=13$~TeV data,''
  CMS-PAS-EXO-16-040.

\bibitem{EXO-16-011} 
  CMS Collaboration,
  ``Search for Dark Matter Produced in Association with a Higgs Boson Decaying to Two Photons,''
  CMS-PAS-EXO-16-011.

\bibitem{EXO-16-012} 
  CMS Collaboration,
  ``Search for dark matter in association with a Higgs boson decaying into a pair of bottom quarks at $\sqrt{s} = 13$~TeV with the CMS detector,''
  CMS-PAS-EXO-16-012.

\bibitem{CMS} 
  CMS Collaboration,
  ``The CMS experiment at the CERN LHC,''
  JINST {\bf 3}, S08004 (2008).


\bibitem{Cacciari:2008gp}
  M.~Cacciari, G.~P.~Salam, and G.~Soyez,
  ``The Anti-k(t) jet clustering algorithm,''
  JHEP {\bf 0804}, 063 (2008).

\bibitem{Catani:1993hr} 
  S.~Catani, Y.~L.~Dokshitzer, M.~H.~Seymour, and B.~R.~Webber,
  ``Longitudinally invariant $K_t$ clustering algorithms for hadron hadron collisions,''
  Nucl.\ Phys.\ B {\bf 406}, 187 (1993).


\bibitem{fastjet}
  M.~Cacciari, G.~P.~Salam, and G.~Soyez,
  ``FastJet User Manual,''
  Eur.\ Phys.\ J.\ C {\bf 72}, 1896 (2012).

\bibitem{Ellis:2009me} 
  S.~D.~Ellis, C.~K.~Vermilion, and J.~R.~Walsh,
  ``Recombination Algorithms and Jet Substructure: Pruning as a Tool for Heavy Particle Searches,''
  Phys.\ Rev.\ D {\bf 81}, 094023 (2010).

\bibitem{Thaler:2010tr} 
  J.~Thaler and K.~Van Tilburg,
  ``Identifying Boosted Objects with N-subjettiness,''
  JHEP {\bf 1103}, 015 (2011).



\bibitem{puppi} 
  D.~Bertolini, P.~Harris, M.~Low, and N.~Tran,
  ``Pileup Per Particle Identification,''
  JHEP {\bf 1410}, 059 (2014).

\bibitem{msd} 
  A.~J.~Larkoski, S.~Marzani, G.~Soyez, and J.~Thaler,
  ``Soft Drop,''
  JHEP {\bf 1405}, 146 (2014).


\end{thebibliography}
\end{document}